\documentstyle[amsfonts,preprint,aps]{revtex}
\tightenlines
\begin{document}
\title{Light propagation control by finite-size effects in photonic crystals}
\author{E. Centeno, D. Felbacq}
\address{LASMEA UMR CNRS 6602\\
Complexe des C\'{e}zeaux\\
63177 Aubi\`{e}re Cedex\\
France}
\maketitle

\begin{abstract}
We exhibit the strong influence on light propagation of the finite size in
photonic band-gap material. We show that light emission can be controlled by
the symmetry group of the boundary of the finite device. These results lead
simply to important practical applications.

{\it Pacs: 42.70 Qs, 11.30-j, 42.82.-m, 03.50.De, 75.40.Mg}
\end{abstract}

\bigskip

Photonic crystals are expected to permit the realization of devices for
integrated optics or laser microcavities \cite{weisbuch,vill,band,JOSA,joan}%
. The today technology now authorizes the making of bidimensional photonic
crystals in the optical range\cite{vikto}. From the theoretical point of
view, there are two ways for characterizing PBG materials. The first way is
the one commonly used in solid state physics where pseudo-periodic boundary
conditions are imposed leading to Bloch waves theory. This is a powerful
tool for the computation of the band structure of the material \cite{bloch}.
However, this theory cannot be used when dealing with the scattering of an
electromagnetic wave by a finite device (which is the actual experimental
situation)\cite{tayeb,nous1,Felbacq}. In that case, boundary effects cannot
be skipped and must be taken into account. In this letter, we address the
problem of the effect on light propagation of both the symmetry group of the
lattice and the symmetry group of the boundary of a finite piece of the
material. The general study of the symmetry of the lattice modes has already
been addressed in order to understand the possible degeneracies and the
existence of full band gaps\cite{cassagne,villeneuve,ville2,shanhui,shanhui2}%
. Here, we show that the finite size of the device strongly modifies the
behaviour of the electromagnetic field, and that the symmetry of the
boundary is a crucial parameter for the control of light emission from PBG
devices.

We deal with a bidimensional photonic crystal that is made of a finite
collection of parallel dielectric rods of circular section. The rods are
settled according to a lattice with some symmetry group ${\bf G}_{Y}$ ($Y$
denotes the elementary cell) of the plane. The relative permittivity of the
lattice is a $Y$-periodic function $\varepsilon _{r}\left( x,y\right) $. The
rods are contained in a domain $\Omega $, having a symmetry group ${\bf G}%
_{\Omega }$. As we deal with objects that are embebbed in an affine
euclidean space, both groups must be given as subgoups of the group of plane
isometries ${\Bbb O}\left( 2\right) $, in a canonical oriented basis. This
is due to the fact that the use of the abstract groups does not permit to
distinguish between two isomorphic realizations. Indeed, using the abstract
groups and unspecified representations of degree 2, we could not, for
instance, distinguish between two squares $C_{1,2}$ deduced from one another
through a rotation $r$ of angle $\pi /4$: denoting $\Gamma _{1}$ a
realization of $D_{4}$ as an invariance group of $C_{1}$, then $\Gamma
_{2}=r\Gamma _{1}r^{-1}$ is a representation of $C_{2}$ and therefore the
equivalence class of $\Gamma _{1}$ is not sufficient to compare the two
figures and a canonical basis has to be precised.\newline
Denoting $1_{\Omega }$ the characteristic function of $\Omega $ (which is
equal to $1$ inside $\Omega $ and $0$ elsewhere), the relative permittivity
is given by $\varepsilon _{\Omega }\left( x,y\right) =1+1_{\Omega }\left(
\varepsilon _{r}\left( x,y\right) -1\right) $. Assuming a $s$-polarized
incident field (electric field parallel to the rods), the total electric
field verifies the d'Alembert equation: 
\begin{equation}
c^{-2}\partial _{tt}E_{z}=\varepsilon _{\Omega }^{-1}\Delta E_{z}
\label{propag}
\end{equation}
Our aim is to study the invariance of this equation under the action of the
various groups characterizing the geometry of the problem. For arbitrary
cross-sections of the fibers, we should also introduce their symmetry group
(for instance the group $D_{4}$ for square rods). The choice of circular
rods simplifies the study in that their symmetry group ${\Bbb O}\left(
2\right) $ contains both ${\bf G}_{Y}$ and ${\bf G}_{\Omega }$. Note however
that strong effects can be obtained by using peculiar symmetries of the
rods, including the enlargment of the band gaps\cite{qiu}. Let us now denote 
$\Gamma \left( {\bf G}_{Y}\right) $ and $\Gamma \left( {\bf G}_{\Omega
}\right) $ the groups of operators associated to ${\bf G}_{Y}$ and ${\bf G}%
_{\Omega }$ respectively\cite{group}. Both operators $\Delta $ and $\partial
_{tt}$ commute with $\Gamma \left( {\bf G}_{Y}\right) $ and $\Gamma \left( 
{\bf G}_{\Omega }\right) $. However, due to the function $\varepsilon
_{\Omega }$, the propagation equation is only invariant under the
intersection group $\Gamma \left( {\bf G}_{Y}\right) {\bf \cap }\Gamma
\left( {\bf G}_{\Omega }\right) $. This simple remark is a crucial point in
understanding the invariant properties of finite crystals and it leads to
extremely important effects in practical applications: it is a clue to
controlling the directions of light propagation in the structure. Indeed,
due the boundary of $\Omega $, the degree of the global symmetry group of
the device is reduced and consequently, from selection rules, the number of
forbidden directions increases.

In order to make this reasoning more explicit, we present two numerical
experiments obtained through a rigorous modal theory of diffraction\cite
{Felbacq,lie} that takes into account all of the multiple scattering between
rods (in case of an infinite lattice this is exactly the KKR method),
moreover it has been succesfully compared with experiments \cite{sabouroux}.
This method allows to deal with finite structure without periodizing tricks 
\cite{mcph} that may lead to spurious phenomena\cite{mozol}. As a lattice
group ${\bf G}_{Y}$, we choose the diedral group $D_{6}$ \cite{eyring}, so
that the lattice has a hexagonal symmetry. The distance between neighbouring
rods is denoted by $d$. In order to create a defect mode in that structure,
we open a microcavity at the centre of the crystal by removing a rod (see 
\cite{tayeb,nous1,villeneuve,figotin,lin} for studies of the properties of
defects in photonic crystals). A defect mode appears within the first gap at
a complex wavelength $\lambda /d=2.264+0.06i$ (using a harmonic
time-dependence of $e^{-i\omega t}$). Such a structure can be used as a
resonator coupled to waveguides\cite{shanhui}.

In the first experiment, we choose the same group for the boundary as that
of the lattice (fig.1)(i.e. ${\bf G}_{\Omega }=D_{6}$). In that case the
propagation equation is completely invariant under $\Gamma \left(
D_{6}\right) $. We plot the map of the Poynting vector modulus associated to
the defect mode (fig.1) and the radiation pattern is given in fig.2. Clearly
the field shows a hexagonal symmetry, which is obvious from the invariance
of the d'Alembert equation. However, when designing light emitting devices,
one whishes to control the direction of light emission. In this example,
there are too many directions of emissions: such a device is to be coupled,
for instance to waveguides, and to get a good transmission ratio, one needs
to concentrate the field in a few useful directions. As it has been stated
above, the number of authorized directions can be reduced by reducing the
global symmetry group $D_{6}$ of the device. This is what we do in the next
numerical experiment where we have changed the boundary so that it has now a
rectangular symmetry (${\bf G}_{\Omega }=\left\{ e,s_{x},s_{y},r\right\} $,
where $s$ denotes a symmetry with respect to $x$ and $y$ respectively and $r$
is a rotation of angle $\pi $), the device is depicted in fig. 3. In that
particular case, the group of the boundary is contained in the group $D_{6}$%
. Then the equation (\ref{propag}) is no longer invariant under $\Gamma
\left( D_{6}\right) $ but solely under $\Gamma \left( D_{6}\right) {\bf \cap 
}\Gamma \left( {\bf G}_{\Omega }\right) =\Gamma \left( {\bf G}_{\Omega
}\right) $ which is strictly contained in $\Gamma \left( D_{6}\right) $. All
the other transformations are now forbidden. That way, we expect a strong
reduction of the directions of propagation of the field.

Indeed, the map of the Poynting vector of the defect mode (fig. 3) as well
as the radiation pattern (fig. 4) shows a strong enhancement of the vertical
direction by forbidding the transverse directions linked to the rotations
and the oblique symmetries. We have designed a resonator that permits to
couple the radiated field in up and down directions with a better
efficiency. 

It should be noted that a group theoretic analysis gives only informations
on the possible directions of emission, the actual directions on which the
field concentrates cannot be obtained by this mean: a rigorous computation
involving a finite structure is then needed. Nevertheless, we have
demonstrated that it was possible to strongly increase the efficiency of
resonators by simply taking into account the symmetry of the boundary of the
device. This remark can be used rather easily in experimental situations and
could lead to a dramatic enhancement of the output of PBG based devices.

\newpage

\bigskip {\bf Figure captions} 

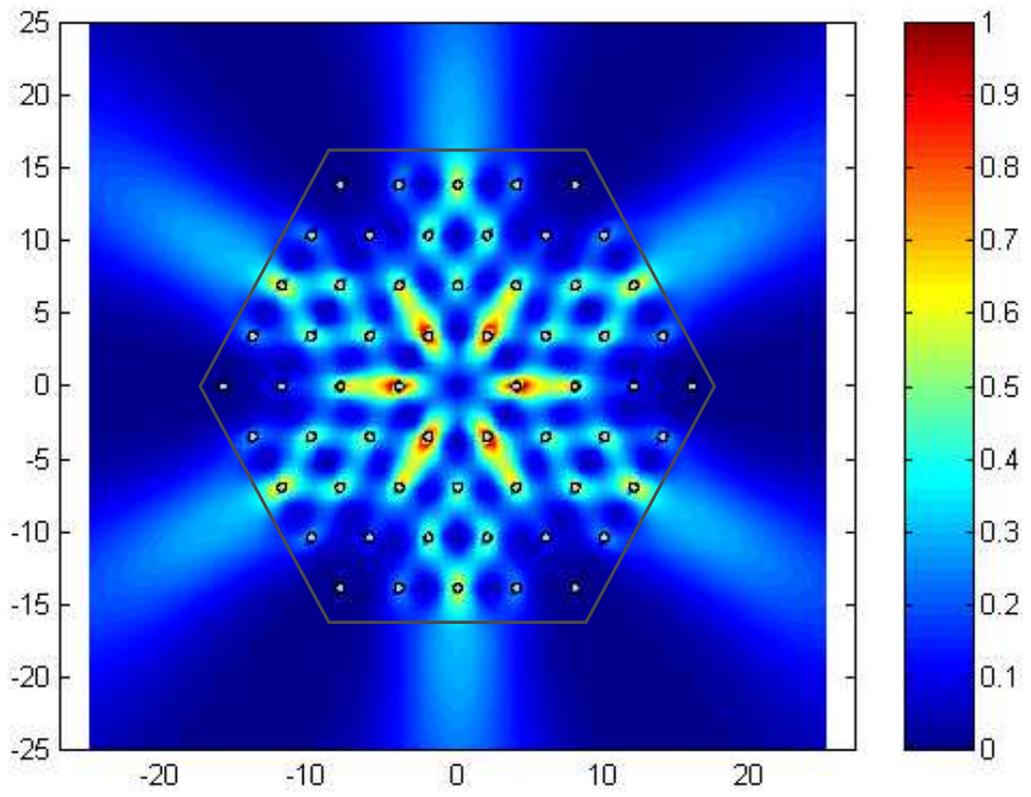
Figure 1: Map of the Poynting vector modulus of the defect mode. Both the
lattice and the boundary have the same symmetry group (${\bf G}_{\Omega
}=D_{6}$). The red line represents the hexagonal symmetry of the boundary of
the crystal. The defect mode possesses all transformations of the hexagonal
point group.\ The ratio of the rod radius to the spatial periode is $r/d=0.15
$ and the optical index is $n=2.9$.

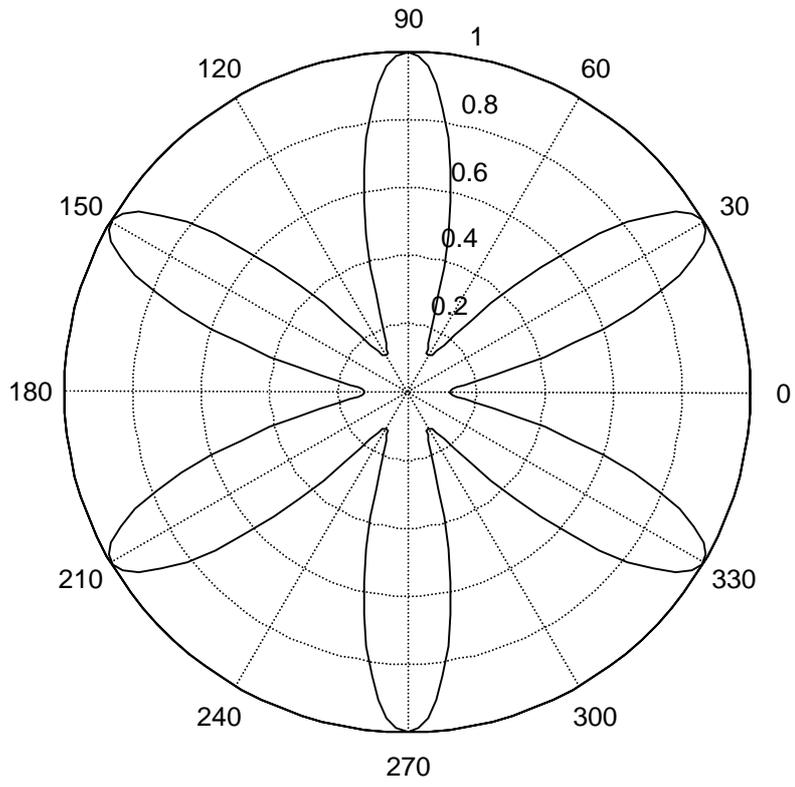
Figure 2: Radiation pattern of the defect mode for the crystal defined in
figure 1. The radiated power is invariant by the hexagonal point group.

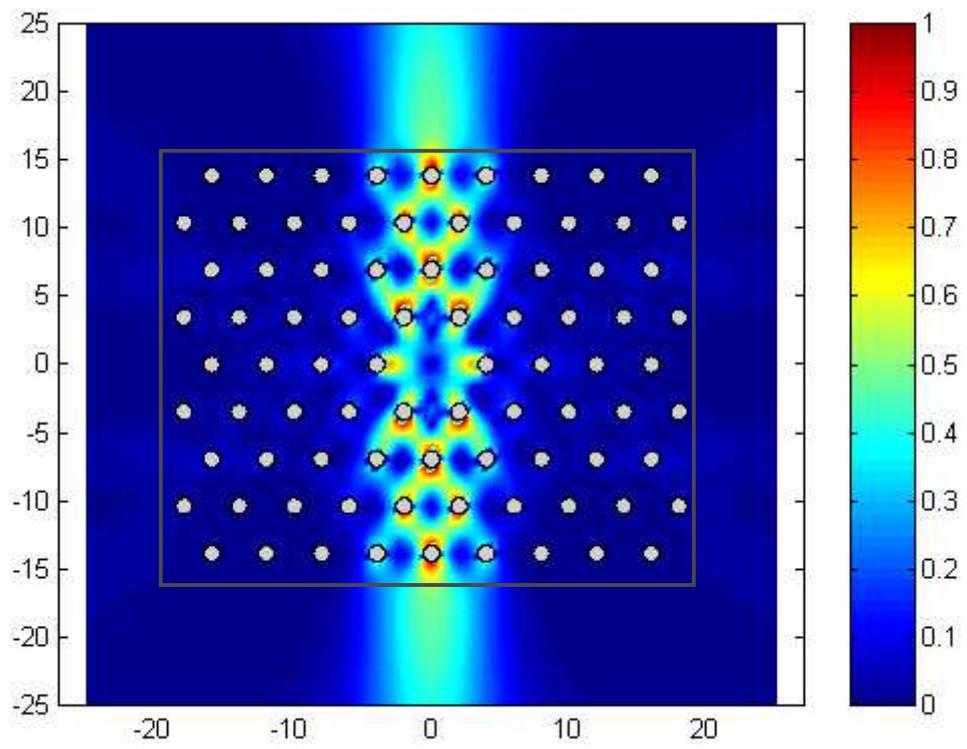
Figure 3: Map of the Poynting vector modulus of the defect mode. The global
symmetry of the crystal is given by the subgroup ${\bf G}_{\Omega }=\left\{
e,s_{x},s_{y},r\right\} $. The red line represents the rectangular symmetry
of the boundary of the crystal. The defect mode is invariant under ${\bf G}%
_{\Omega }$. The ratio of the fiber radius to the spatial periode is $%
r/d=0.15$ and the optical index is $n=2.9$.

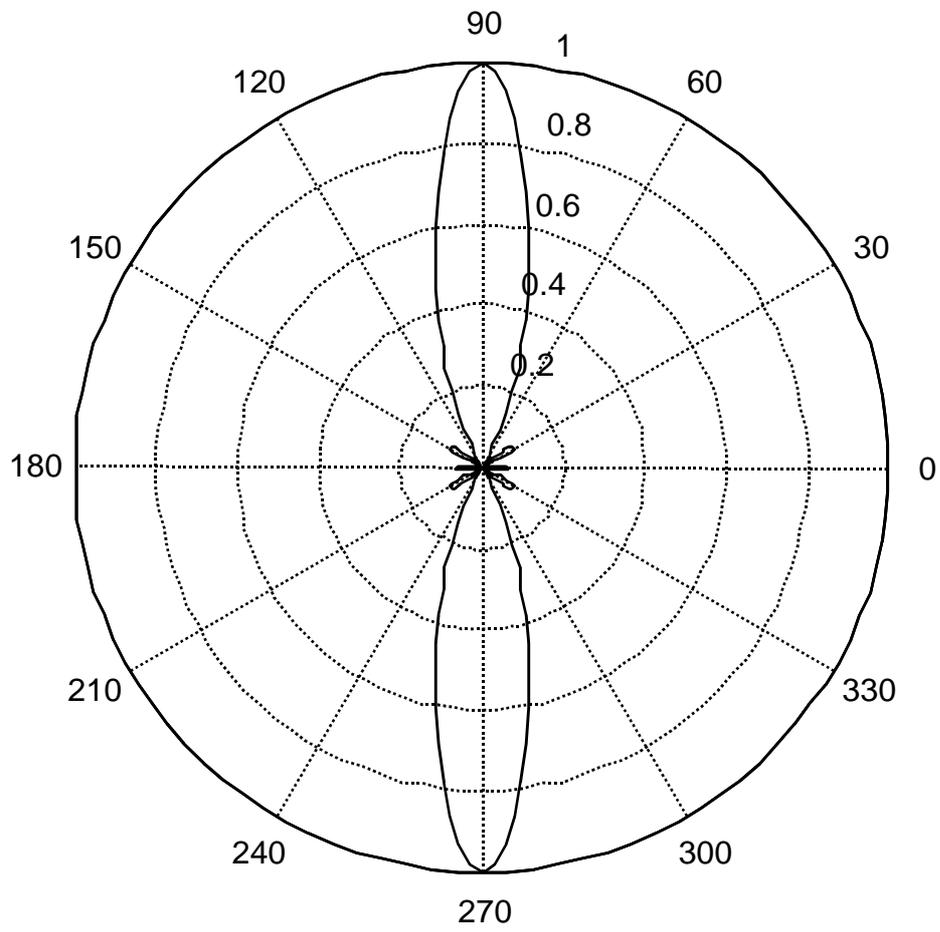
Figure 4: Radiation pattern of the defect mode for the crystal defined in
figure 3. The radiated power is invariant by the subgroup ${\bf G}_{\Omega }$%
.

\end{document}